\documentstyle[epsfig,times,12pt]{article}
\begin{document}
\title{SCIENCE: A Solid Whole}
\author{Bhag C. Chauhan{\footnote{chauhan@cfif.ist.utl.pt}}\\
\it Centro de F\'{i}sica das Interac\c{c}\~{o}es Fundamentais (CFIF)\\
Departmento de Fisica, Instituto Superior T\'{e}cnico\\
Av. Rovisco Pais, 1049-001, Lisboa-PORTUGAL}
\maketitle
\begin{abstract}
However, the observations encompassed by classical physics excludes the observer from the physical reality, yet the deep-down understandung of nature --{\it the quantum theory}-- can not avoid the intrusion of observer into the measurement process. Indeed, the quantum physics experiments have knocked the door of a new paradigm: in which science of consciousness is an important axiom. In the present work, it is argued --by taking into account of the views of learned scientists and philosophers-- that modern science is incomplete and lacking something in the basic understanding of nature \cite{r}. Classical physics has failed to explain the dynamics of the microscopic particles, eventhough modern scientific researches are based upon the prejudice posed by classical physics: keeping the outer physical universe as a separate entity, that is something quite independent of the observer --human mind. One should not forget that human-being is a part of nature and human mind is an essential component of our observations. Basically, it is the observer --the consciousness-- which makes perception possible. The working of human mind must be included in our scientific theories. In fact, human free will is an illusion and nothing in the entire universe (with life) lies outside the domain of science and determinism.   
\end{abstract}
\newpage
\section{Human Perception}
All the physical objects seem to have finite dimensions and some other measurable properties. The human body is physical material having a limited ability of perception of the outside material world. It has five senses of perception: vision, hearing, taste, smell and touch. To make perception possible, the ``electromagnetic-wave'' connects the observables to the observer. Nature has granted a narrow band of the electromagnetic spectrum to our body for perception: a visible band of light and a part  of the infra-red region. The window is being broadened by our scientific knowledge: the scientifically detected (known) part of the spectrum extends from a low frequency of power induction to a high frequencies of cosmic rays ($\lambda~\approx~10^{5}~cm~-~10^{-13}~cm$). We are not aware of, what lies beyond on either side of the band. 
\par
It seems that the electromagnetic spectrum is ``eternal and infinite'', and nobody can deny its infinite extensions on the both sides of the known part --large wavelength region and small wavelength region. The scientific knowledge is just uncovering a region of its infinite extensions, step-by-step, and consequently, understanding nature better. It is only the total knowledge of the infinite spectrum which can touch the ``ultimate truth'' sitting at the heart of nature. The understanding and uncovering of this spectrum up to any finite extension will always be infinitely small as compaired to the infinite whole of it. In this way the efforts of understanding the whole extension of this eternal spectrum, and thereby, to gain the knowledge of universe with life and entire existence are never going to be fulfilled. 
\par
The human body --a frame with the five sense organs and a brain-- is called by some philosophers as ``a limited instrument,'' because of its above discussed limitations. So, doing science with such a blemish instrument, one can never discover the infinite whole; the ``ultimate reality''. For the systems like weather, earthquakes, rolling dices etc... and human behaviour, of course, it has proved scientifically impossible to describe a state of system accurately for a long time into the future. Instead, probabilities can be derived to describe a state which might happen in future. Science is ``unpredictable'' \cite{r3}. The claim of determinism is a failure in science \cite{r2}. In quantum mechanics, the unpredictability of the science has been accepted by postulating {\it ``principle of uncertainty''}. Albert Einstein, a pioneer scientist of the last century, bothered much with this principle and said: ``God does not play dice''. According to this principle, it is impossible to measure every physical variable of a microscopic particle with full accuracy. And the experimental results have proved it. So, recognizing ``uncertainty'' as a principle, has infered the inability of science to discover the ultimate reality; which is ``absolute'' and demands the exactness/ perfectness of measurements. 
\par
No doubt, the main goal of modern science is to discover the ultimate truth and develop a {\it ``Theory Of Everything (TOE)''} but there is an essential question, at this difficult stage of science, which every wise man should ask: ``Can science explain everything?''... and I have a beautiful answer of a learned scientist --Steven Weinberg \cite{r1}: ``Clearly not! There certainly always will be accidents that no one will explain, not because they could not be explained if we knew all the precise conditions that led up to them, but because we NEVER will know all these conditions''. It would be better to add here some comments of Albert Einstein, in his old age: ``I used to think when I was young that sooner or later all the mysteries of existence would be solved, and I worked hard. But now I can say that the more we know, the more our existence turns out to be mysterious. The more we know, the less we know and the more we become aware of the vastness... Science has failed in de-mystifying existence, on the contrary it has mystified things even more''. It seems that as we are extending the radius of our scientific knowledge, so too have been increasing the circumference of our ignorance.

\section{Superstitions and Superpositions}
Superstitions have been developed through the personal experiences, observations, and beliefs of the people in the olden times. Interestingly, most of them survived for a long time because they worked. If we think rationally and review  all the superstitions without any prejudice, we can say that there have been always a probability for each event to happen in a certain way: according to the beliefs of people. This probability have been found higher if a larger number of observers (people) supported happening of the event in that way because there existed a larger confidence level of the observers. In case the event occured in another, way not according to their wishes, they found other solution --make happy their gods-- and their wishes have been found to be fulfilled. In other words, finally, the event happened according to their beliefs and wishes. Certainly, this had been something very important to do with the state-of-mind of the people. In fact, the superstition doctrine is still working well among more than 50\% people of the globe: through many religions, sects and other practices etc... Predictions of a person of true heart with strong beliefs; strong will power; elation and a sidh-purush have been found to be true most of the times. There is a famous saying that if you pray true to the heart, God will listen you.  
\par
Quantum theory is a new paradigm in search of the ultimate reality but playing with the old dices of superstitious human civilizations. The only difference is, the more objectivity in the predictions of the events with a beautiful use of mathematics. In fact, there is no mechanism in theory to select a particular actuality from the multiplicity of superposed possibilities. The superposition, of all the possible states of system, is supposed to represent the final possibility of the event which can appear as a reality. This is nicely explained by the tools of mathematics and the axioms of probability theory. It might have led us further near to the reality but certainly not going to connect with the absolute truth, as the results are probabilistic --a matter of chance-- again. We are again at a predicament situation. One cannot say for sure about the final state of the system; then, what we have gained is just a mathematical cleverness to convince ourselves that the mathematics supports the happening of a particular event with such and such chances. Yet, it is not necessary that any sets of laws which are completely consistent mathematically, can surely describe nature as we observe it \cite{r1}. 
\par
There has been a paradigm shift from superstitions of human beliefs to superpositions of quantum mechanics; from an abstractness to the concrete mathematical equations; from a matter of luck to a matter of chance. The superstitions have been found based on a belief that a supernatural power works, and the superposition is based on a belief that the mathematical power works. In superstitions, one have has no mean how to explain; only the personal belief was enough to tell people and it worked well in that period of history and still working well among most of the common mass. On the other hand, about superposition/ probability theory one can write a book, one can furnish a library; because there is something concrete to show. Now, the reality is being accepted as an objective thing, which can be explained on a plane paper; it is nothing to do with the subjective experience. But, either one of the two is not a complete in itself. They are totally opposite poles of the same magnet (one should say), which certainly has a joint at center. We are talking either about its north pole or about its south pole, which alone does not constitute a total whole of it. Basically, truth does not exist at either of the two poles (superstition or superposition) but at the center; at the union of two pseudo-realities. Now, another paradigm shift is essential; which can unite and dissolve these two schools of thoughts. Indeed, the quantum physics experiments have knocked the door of a new paradigm: At the union, both the apparent realities dissolve into the ultimate one.       

\section{Cosmology: An Interface}
 As far as the origin and evolution of this universe --cosmology-- is concerned, the most popular theory among the scientists is the ``big bang theory''. Although this theory has passed some scientific tests, there are still many more trials which it must undergo successfully. Scientific evidence strongly supports that the universe had a definite beginning a finite amount of time ago. According the big bang theory there was nothing before the big bang and all the space-time must have originated there and then. No matter/energy could exist before this bang as there was no space and time for it to be in. In the context of a recent test of this theory, John Bahcall (a learned astrophysicist) says \cite{r6}: ``I am happy that the big bang theory passed this test, but it would have been more exciting if the theory had failed and we had to start looking for a new model of the evolution of universe''. In fact, there are many domains of mainstream cosmology that are far from being settled. 
\par
The biggest problem of the big bang theory of the origin of the universe is philosophical --perhaps even theological-- what was there before the bang? \cite{r5}. This stands as an embarrassing situation for the scientists. Robert Jastrow --the first chairman of NASA's Lunar Exploration Committee-- himself admitted \cite{r5a}: ``Astonomers try not to be influenced by philosophical considerations. However, the idea of a universe that has both a beginning and an end is distasteful to the scientific mind''. 
\par
To avoid this initial difficulty the idea of {\it singularity} was introduced in which the universe expands from a singular point (a point of infinite density at which the laws of physics break down) and collapses back to the singular point and repeats the cycle indefinitely \cite{r5}. The idea was appreciated to avoid the theistic base of the theory but the experimental evidence seems to indicate that this type of oscillating universe is a physical impossibility and facts supports that the universe will expand for ever \cite{r8}. The attempts behind this idea to avoid theistic beginning of the universe all fail \cite{r9}. The philosophical origin of the big bang cannot be denied.
\par
It has been known for some time that the fundamental constants of physics (particle masses, coupling constants of the various forces etc...) are within a narrow range and fine-tunned for life to exist. In the words of Steven Weinberg \cite{r4}: ``It is almost irresistible for humans to believe that we have some special relation to the universe, that human life is not just a more-or-less farcial outcome of the chain of accidents reaching back to the first three minutes, but that we were somehow built in from beginning''. Indeed, the reality that physics addresses is only a part of the total interacting reality which holds the key of the philosophical origin of big bang theory. Arthur S. Eddington --a brilliant astronomer-- once said: ``The universe no longer looks like a thing but like a thought''. In fact, modern cosmology has an interface with a {\it subtle science}: the science of consciousness. 

\section{Modern Science \& Human Consciousness}
However, the observations encompassed by classical physics excludes the observer from the physical reality; yet, the deep-down understandung of nature --{\it the quantum theory}-- can not avoid the intrusion of observer into the measurement process. 
It is well acknowledged that the classical physics is an incomplete understanding of nature. Eventhough, our scientific researches are based upon the prejudice posed by classical physics: keeping the outer physical universe as a separate entity, that is something quite independent of the observer.
A serious flaw is existing in the grass-root level working of science \cite{r}. It will be wise to understand fully the working of this objective method of scientific studies which is only relying upon the limited sense perception of human body. Basically , we are separating the real observer from the observation. To make a perception possible, there must be a subject --consciousness-- who can observe a phenomenon or an event with the help of a connecting principle. It is not the physical part of human brain which acts as the observer (the knower) and make the perception possible but there is a subtle playback entity {\it ``Mind''}, which we donot consider in our scientific theories. The human mind is the doer, the observer which interprets the messages collected from outside by brain with the help of sense organs and instruments. It makes a scientist to recognize or refute the existence of an object or a phenomenon. ``Nothing in current science can account for consciousness, yet consciousness is the one thing we can not deny'': a revolutionary futurist, Peter Russell, says \cite{r4a}.
\par
According to Eugene Wigner --American Physicist, nobel laureate: ``The next revolution in physics will occur when the properties of mind will be included in the equations of quantum theory''. The most creative physicists have always emphasized that human consciousness is at the foundation of the scientific method behind physics \cite{r7}. Erwin Schrodinger --one of the founders of quantum mechanics-- felt deeply that human mind is a sole constructor of all the observations and quoted as: ``Our picture of the world is, and always will be, a construct of the mind''. This is the biggest challenge for the scientists, namely to include the human consciousness in their theories. However, it can be proved hard for the scientific community to consider the great importance of human element in doing science but it's worth-doing and enjoyable. Science is ``fun'' \cite{r3}. The fun starts happening in doing science or any work of day-to-day life, when one accepts the full co-operation of his inner-being. And the climax of fun can be achieved when one accepts this inner-being as an essential part of scientific observations and practice it in the objective world (real life). 
\par
A man who has developed a big scientific panorama, it is unbelievable that he himself is non-scientific. Never! He is scientific from his inner-being and only then he could discover science which is always there, the ultimate. Study of inner-being is also called  ``spirituality''. Spirituality is a pure form of religion --free from sectarian, orthodox and dogmatic idealogy. This is not an utopian theory but a scientific and practical philosophy which can be practised and realised in day-to-day life. A study in the latest edition of ``The Medical Journal of Australia'' asserts that those who consider the spiritual dimension essential to their lives not only live longer, they are also healthier --with lower blood pressure, lower cholesterol and lower rates of some cancers-- and less likely to abuse drugs and alcohol. More than 500 scientific studies conducted at 200 independent universities and institutions in 33 countries and published in over 100 leading scientific journals have been documented, which benefits every sphere of life: physiological, psychological, sociological, and ecological \cite{r7}. 

\section{Determinism and Free Will}
``Science is determinist; it is so {\it a piori}; it postulates determinism, because without this postulate science could not exist'', stated by Henri Poincare -- a great mathematician and physicist. If there is a scientific law working anywhere in the universe, determinism must be there. According to Pierre Simon Laplace --a French astronomer and mathematician-- the entire future course of the universe is laid out as a consequence of two deterministic factors: 1) the laws of nature, and 2) the state of universe at any moment of time.
Universe is nothing but a collection of numerous phenomena. Anything happening around us, in the universe, is controlled by a certain set of physical laws. These laws determine the birth, life and death of all the events of universe. So, with the total understanding of these physical laws one can predict the future of an event very well in advance. This is the beauty of science, as Henri Poincare quoted. Everything is deterministic and there is no place for a ``free will'' to exist. 
\par
The principle of determinism may disappear for some time because of our incomplete understanding of the scientific laws; yet, it does not imply that the free will is prevailing. There must be certain hidden variables which are defusing the realisation of the principle of determinism.
The conscious mind receives information from the physical world only through eyes and other sense organs that are presumably understandable deterministically. Determinism also implies the objectivity of the phenomena. In principle, there can be nothing subjective in the entire existence. Even the true observer --human mind-- is objective in nature. It keeps subjectivity only at its superficial layers. However, the deep-down part of it, is purely objective in nature. This shallow subjectivity is a natural consequence of the inherit genetic codes (informations recorded through the long process of evolution) and the influence of environment; which can be scientifically decoded and understood. So, in this way, human behaviour --which we suppose as the most complex system-- is also deterministic. In fact, everything in the entire existence is objective and within the domain of science and the principle of determinism is prevalent everywhere. 
\par
A free will can exist only if there is not at all anything else existing in the entire whole. Only a source totally detached from nature (matter) which is the origin and cause of everything --options, thoughts, feelings, etc... that is the existence of an ``autonomous mind'', i.e. a {\it principium individuationis} \cite{r7a}... the entity which is {\it ``Only One and Supereme''} has the privilege to enjoy the glory of free will. As soon as the {\it``One becomes Two''}, a boundary appears, laws hold and the free will concept is lost; as now there exists another entity which introduced the physical laws in order to justify its identity-- {\it as the number ``Two''}. As a consequence, the free will is automatically disappeared. As soon as the {\it``One becomes Many''}, of coarse, the more complex laws hold and the individual conscious will becomes more constrained. Now the will of the every conscious object is guided by the resultant force of all the natural laws. 
\par
Universe is at the stage of {\it``One to Many''}, where the complex physical laws hold and the determinism is prevailing everywhere. The free will seems to be prevailing in human behaviour, but this is an illusion. There exists a will, which is not free but constrained -- a ``constrained will''. Basically, there is always a window of priorities with in which the human-will has to work. This conscious will is manifest through deterministically understandable science. All the thoughts, values, volitions, decisions, acts, are the product of physical, chemical, or physiological and psychological processes going on in the human body. Our will is a result of our past life experiences, the innate genetic codes which have been acquired through the struggle of long time in the process of evolution and the present physiological and psychological states of our body. By the way, it is an essential feature of the constrained will, which makes life interesting and struggleful. A life with a free will is not at all, logically possible. But if somehow it happened to be so, can never be worth-living but all boring and suicidal. 

\section{Conclusions}
The observer problem can be evaded in classical physics by excluding the consideration of the mind of observer but this option is certainly unavailable for quantum mechanics. The intrusion problem of observer occurs at the level of the quantum mechanical experiment clearly shows that modern science is incomplete understanding of nature. Our scientific researches are based upon the prejudice posed by classical physics: keeping the outer physical universe as a separate entity, that is something quite independent of the observer --human mind. One should not forget that human-being is a part of nature and human mind is an essential component of our observations. Indeed, it is the observer --the knower-- which makes perception possible. The working of human mind must be included in our scientific theories. Moreover, limits of human perception (excluding mind) and philosophical base of modern cosmology have disclosed another physical evidence for the existence of a bigger reality and a new paradigm which can encompass the human consciousness and solve all the difficulties that modern science is waging with.
\par
In practice, human behaviour appears un-deterministic and the free will seems to be working there. But this is our ignorance about the scientific laws that holds the key of the subtle dynamics of our inner-being. Basically, there are two kinds of sciences: one, the objective science --{\it ``modern science''}-- that studies the outer --material nature; and two: subjective science, a study of inner-being, super nature --{\it ``science of consciousness''}. If both are joined for completeness, only ``SCIENCE'' remains as the ultimate knowledge (wisdom); which can explain everything. Scientist should accept the challenge and start searching for the possibilities to include the human consciousness in the observations and theory. Only this path could explore the real potential of the human being and create a man in the plenitude of wisdom; who can solve the mystery of quantum theory, and thereby, reabilitate the principle of determinism. In principle, that would fulfil his candidature to uncover the vastness of the electromagnetic spectrum and enjoy the ecstasy of the ultimate reality. So, it is only the completeness of science --``SCIENCE''-- which is capable of uttering the ultimate answer to the human quest: enabling man to develop the ``TOE'' and thereby, bringing upon him benign grace of the ultimate truth. 

\end{document}